\documentclass[final, sort&compress]{elsarticle}
\usepackage[top=1in, bottom=0.7in, left=1in, right=1 in]{geometry}
\usepackage{amsmath,amssymb,amsthm}
\usepackage{graphicx} 
\usepackage{bm}       
\usepackage{mathtools}
\usepackage{dcolumn}  
\usepackage[colorlinks=true,linkcolor=blue]{hyperref} 
\usepackage{booktabs} 
\usepackage[mathscr]{euscript} 
\usepackage{ulem}     
\usepackage{xcolor}   
\usepackage{natbib}   

\newcommand{\bea}{\begin{align}}
\newcommand{\eea}{\end{align}}
\newcommand{\beq}{\begin{equation}}
\newcommand{\eeq}{\end{equation}}

\begin{document}
		\title{
	Analyzing the relationship between infinite symmetries and $N$-soliton solutions in the AKNS system	
	}	
		\author[rvt1]{Xiazhi Hao}%
		\author[rvt2,rvt3]{S. Y. Lou\corref{cor1}}
		\cortext[cor1]{Corresponding author. Email: lousenyue@nbu.edu.cn(S. Y. Lou) }
		
		\address[rvt1]{School of Mathematical Sciences, Zhejiang University of Technology, Hangzhou, 310014, China}
		\address[rvt2]{School of Physical Science and Technology, Ningbo University, Ningbo, 315211, China}
		\address[rvt3]{Institute of Fundamental Physics and Quantum Technology, Ningbo University, Ningbo, 315211, China}

		\begin{abstract}	
			This paper investigates the algebraic reduction of the infinite-dimensional symmetries of the Ablowitz–Kaup–Newell–Segur system when restricted to multi-soliton solution. By systematically analyzing, we demonstrate that the entire $K$-symmetry hierarchy collapses into a finite-dimensional module over the field of wave parameters, spanned by elementary center-translation generators. Higher order $K$-symmetries are explicitly reconstructed as linear combinations of these basis vectors.
			In contrast, $\tau$-symmetries resist such decomposition on pure soliton backgrounds, however, they become finite-dimensional once the underlying solution space is extended to the full multi-wave manifold, which carries more independent wave parameters. We construct an explicit basis consisting of four fundamental symmetry vector fields, two lowest $K$-symmetries and two primary $\tau$-symmetries, in terms of which all higher $\tau$-symmetries are uniquely expressible as linear combinations of these symmetry vector fields.
			These findings not only clarify the algebraic structure of infinite symmetries on special solution, but also provide an algorithmic framework for deriving exact multi-wave solutions of integrable systems.
		\end{abstract}

		\maketitle
		\large
		\section{Introduction}
Integrability \cite{int1,abib10}, once perceived as a mathematical curiosity, has quietly become the organizing principle behind some of the most robust wave phenomena observed in nature. 
Contemporary research has further revealed deep connections between integrability and fundamental field theories through twistor geometry \cite{twist} and self-dual Yang-Mills constructions \cite{yang, yang1}. Particularly compelling is the Ward hypothesis \cite{ward}, which posits that all integrable systems may emerge as dimensional reductions of self-dual Yang-Mills theory - a conjecture verified for key models including the KdV, sine-Gordon, and nonlinear Schr\"odinger equations. Within this unified framework, the Ablowitz-Kaup-Newell-Segur (AKNS) scheme \cite{akns74,cbib8,bbib9,wbib5,sbib45,Cardenas2021} occupies a privileged position as a master system that not only incorporates these celebrated (1+1)-dimensional models but also generates novel physical applications across multiple domains.
		
The AKNS system, characterized by its coupled nonlinear Schr\"odinger equations \cite{akns74,wbib29}
		\begin{eqnarray}\label{akns}
		\begin{pmatrix}
		q \vspace{0.2cm}\\
		r 
		\end{pmatrix}_t={\rm i}\begin{pmatrix}
		q_{xx}-2q^2r \vspace{0.2cm}\\
		-r_{xx}+2r^2q 
		\end{pmatrix},
		\end{eqnarray}
has been extensively studied for its ability to generate soliton solutions, which are self-reinforcing wave packets that exhibit particle-like stability and maintain their shape and speed upon interaction. The soliton solutions of the AKNS system are not only mathematically intriguing but also possess profound physical relevance, as they model various wave phenomena in nature and technology. 
These particle-like pulses shepherd femtosecond light through kilometer-long fibers \cite{fiber}, sculpt coherent atom clouds in Bose-Einstein condensates \cite{be,be1}, and, most speculatively, ripple across curved space-time as exotic gravitational solitons, proving that one compact integrable equation can speak for optics, matter waves and gravity alike \cite{hbib15,op}.

Central to the integrability of the AKNS system are its infinite symmetries, which encode the conservation laws and dynamical invariants. Notably, the \(K\)-symmetries, generated by the recursion operator \(L\), form an infinite dimensional algebra that underpins the solvability and multi-soliton dynamics. While these symmetries are well-studied mathematically, their physical interpretations and interdependencies in the context of soliton solutions remain under explored \cite{lou1,mbib26,25pla,jl1}. For instance, how do higher order \(K\)-symmetries relate to the geometric and kinematic properties of multi-soliton solutions? Do these symmetries constrain soliton interactions or reveal hidden physical constraints?  
	
In this work, we address these questions by systematically analyzing the relationships among \(K\)-symmetries, \(\tau\)-symmetries for exact \(N\)-soliton solutions of the AKNS system.

\section{Soliton solutions of the AKNS system}

As the fundamental integrable reduction of the Zakharov–Shabat spectral problem, the AKNS system supplies a complete, algebraically explicit framework for multi-soliton phenomena. Its \(N\)-soliton manifold, parametrized by reflectionless scattering data, constitutes an exact nonlinear superposition of localized waveforms. In what follows we condense this nonlinear superposition into a markedly more compact and structurally transparent representation, thereby delivering a refined formula for the \(N\)-soliton solution \cite{sbib54,xbib15,mbib27} that is both computationally expedient and amenable to systematic asymptotic analysis.
Below we record two closed-form representations of the $N$-soliton solution of the AKNS system, each exposing a distinct facet of its integrable structure.

\subsection{Type-I representation: N-soliton solution via rational function structure}

	The first representation is realized through a rational fraction.
	Define the phase variable
\begin{eqnarray*}
\xi_i=k_ix-\mathrm{i}k_i^2t+\xi_{i0},
\end{eqnarray*}
then the \(N\)-soliton solution of the AKNS system reads
\begin{eqnarray}\label{fnsolution}
&&q_n=\frac{Q_n}{M_n},\quad r_n=-\frac{R_n}{M_n}
\end{eqnarray}
with numerator and denominator constructed by partial summations over index sets  
\begin{small}
\begin{eqnarray*}
	R_n&=&  \prod_{i=1}^n k_i^2 \prod_{1 \leq i < j \leq n} (k_i - k_j)^2 \exp\left({\sum_{i=1}^n \xi_i}\right), n\geq 2,\\ 
	Q_n&=&\sum_{\substack{I \subseteq \{1,\dots,n\} \\ |I|=n-2}} \left(\prod_{\substack{i\in I}}k_i^2 \prod_{\substack{i,j \in I \\ i < j}} (k_i-k_j)^2  \exp\left( \sum_{k \in I} \xi_k \right)\right)+\sum_{\substack{I \subseteq \{1,\dots,n\} \\ |I|=n-1}}\prod_{1 \leq i < j \leq n}^{} (k_i - k_j)^2 \exp\left({\sum_{i=1}^n \xi_i}\right), n\geq 3,\\
	M_n&=&\sum_{\substack{I \subseteq \{1,\dots,n\} \\ |I|=n-1}} \left(\prod_{\substack{i\in I}}k_i^2 \prod_{\substack{i,j \in I \\ i < j}} (k_i-k_j)^2  \exp\left( \sum_{k \in I} \xi_k \right)\right)+\prod_{1 \leq i < j \leq n} (k_i - k_j)^2 \exp\left({\sum_{i=1}^n \xi_i}\right), n\geq 3.
\end{eqnarray*}
\end{small}
For completeness we list the lowest-index cases. The one-soliton reduces to
\begin{eqnarray}\label{1-sol}
&&q_1=\frac{1}{1+{\rm e^{\xi_1}}}, \quad r_1=-\frac{k_1^2{\rm e^{\xi_1}}}{1+{\rm e^{\xi_1}}}, \quad \xi_1=k_1x-{\rm i}k_1^2t +\xi_{10},
\end{eqnarray}
while the two-soliton solution (\(n=2\)) is explicitly
\begin{eqnarray}\begin{split}\label{2-sol}
&&q_2=\frac{1+{\rm e^{\xi_1}}+{\rm e^{\xi_2}}}{k_1^2{\rm e^{\xi_1}}+k_2^2{\rm e^{\xi_2}}+(k_1-k_2)^2{\rm e^{\xi_1+\xi_2}}},\\
&&r_2=-\frac{k_1^2k_2^2(k_1-k_2)^2{\rm e^{\xi_1+\xi_2}}}{k_1^2{\rm e^{\xi_1}}+k_2^2{\rm e^{\xi_2}}+(k_1-k_2)^2{\rm e^{\xi_1+\xi_2}}}
\end{split}
\end{eqnarray}
with $\xi_1=k_1x-{\rm i}k_1^2t +\xi_{10},~\xi_2=k_2x-{\rm i}k_2^2t +\xi_{20}$.

The three-soliton case (\(n=3\)) is obtained by substituting the corresponding expressions \begin{eqnarray*}
	Q_3&=&k_1^2{\rm e^{\xi_1}}+k_2^2{\rm e^{\xi_2}}+k_3^2{\rm e^{\xi_3}}+(k_1-k_2)^2{\rm e^{\xi_1+\xi_2}}+(k_1-k_3)^2{\rm e^{\xi_1+\xi_3}}+(k_2-k_3)^2{\rm e^{\xi_2+\xi_3}},\\
	R_3&=&k_1^2k_2^2k_3^2(k_1-k_2)^2(k_1-k_3)^2(k_2-k_3)^2{\rm e^{\xi_1+\xi_2+\xi_3}},\\
	M_3&=&k_1^2k_2^2(k_1-k_2)^2{\rm e^{\xi_1+\xi_2}}+k_1^2k_3^2(k_1-k_3)^2{\rm e^{\xi_1+\xi_3}}+k_2^2k_3^2(k_2-k_3)^2{\rm e^{\xi_2+\xi_3}}\\&+&(k_1-k_2)^2(k_1-k_3)^2(k_2-k_3)^2{\rm e^{\xi_1+\xi_2+\xi_3}}
\end{eqnarray*}  into the general quotient 
\begin{eqnarray*}
&&q_3=\frac{Q_3}{M_3}, \quad r_3=-\frac{R_3}{M_3}
\end{eqnarray*}
with $\xi_i=k_ix-{\rm i}k_i^2t +\xi_{i0},~i=1,2,3$.

This representation makes manifest the pairwise interaction of solitons through the differences \((k_i-k_j)\) and is particularly convenient for analysing the large-time separation.

\subsection{Type-II representation: N-soliton solution via Hirota bilinear formalism }
The second representation relies on the Hirota bilinear formalism and expresses the fields as ratios of multivariate exponential polynomials.  Introduce auxiliary phases 
\begin{eqnarray*}
	 \xi_i=k_ix+{\rm i}k_i^2t+\xi_{i0},\quad  \eta_i=l_ix-{\rm i}l_i^2t+\eta_{i0},
\end{eqnarray*}
the $n$-soliton solution \cite{mbib27} is then expressed as
\begin{eqnarray}\label{nsoliton}
q_n=\frac{g_n}{f_n},\quad r_n=\frac{h_n}{f_n}
\end{eqnarray}
through a set of exponential polynomials, where the coefficients are derived from the interaction parameters of the solitons. Specially, the functions $g_n, h_n$ and $f_n$ are defined as follows	
\begin{eqnarray*}
&&g_n = \sum_{\substack{I,J \subseteq \{1,\dots,n\} \\ |I|=|J|+1}} C_{I,J} \exp\left( \sum_{i \in I} \xi_i + \sum_{j \in J} \eta_j \right), \\
&&h_n = \sum_{\substack{J,I \subseteq \{1,\dots,n\} \\ |J|=|I|+1}} D_{J,I} \exp\left( \sum_{j \in J} \eta_j + \sum_{i \in I} \xi_i \right), \\
&&f_n = 1 + \sum_{\substack{I,J \subseteq \{1,\dots,n\} \\ |I|=|J|}} E_{I,J} \exp\left( \sum_{i \in I} \xi_i + \sum_{j \in J} \eta_j \right).
\end{eqnarray*}		
The coefficients \(C_{I,J}, D_{J,I}\) and \(E_{I,J}\) are compactly written as products over all interacting pairs
\begin{eqnarray*}
C_{I,J} &= \left( \prod_{\substack{i_1,i_2 \in I \\ i_1 < i_2}} a_{i_1 i_2} \right) \left( \prod_{\substack{i \in I \\ j \in J}} a_{i j} \right) \left( \prod_{\substack{j_1,j_2 \in J \\ j_1 < j_2}} a_{j_1 j_2} \right), \\
D_{J,I} &= \left( \prod_{\substack{j_1,j_2 \in J \\ j_1 < j_2}} a_{j_1 j_2} \right) \left( \prod_{\substack{j \in J \\ i \in I}} a_{i j} \right) \left( \prod_{\substack{i_1,i_2 \in I \\ i_1 < i_2}} a_{i_1 i_2} \right), \\
E_{I,J} &= \left( \prod_{\substack{i_1,i_2 \in I \\ i_1 < i_2}} a_{i_1 i_2} \right) \left( \prod_{\substack{i \in I \\ j \in J}} a_{i j} \right) \left( \prod_{\substack{j_1,j_2 \in J \\ j_1 < j_2}} a_{j_1 j_2} \right)
\end{eqnarray*}
with 
\begin{eqnarray*}
	&&a_{i_1 i_2}=-(k_{i_1}-k_{i_2})^2, ~a_{j_1 j_2}=-(l_{j_1}-l_{j_2})^2, ~a_{i j}=-\frac{1}{(k_{i}+l_{j})^2}.
\end{eqnarray*}		
For the	one-soliton solution, we have
\begin{eqnarray}\label{1-s}
\begin{split}
&g_1={\rm e}^{\xi_1},~ h_1={\rm e}^{\eta_1},~ f_1=1+a_{12}{\rm e}^{\xi_1+\eta_1},~ a_{12}=-\frac{1}{(k_1+l_1)^2},\\& \xi_1=k_1x+{\rm i}k_1^2t+\xi_{10},~\eta_1=l_1x-{\rm i}l_1^2t+\eta_{10}.
\end{split}
\end{eqnarray}
Explicit expression for \(n=2\) is obtained by extending the above summation rules 

	\begin{eqnarray}\label{2-s}
	\begin{split}
	g_2&={\rm e}^{\xi_1}+{\rm e}^{\xi_2}+a_{12}a_{13}a_{23}{\rm e}^{\xi_1+\xi_2+\eta_1}+a_{12}a_{14}a_{24}{\rm e}^{\xi_1+\xi_2+\eta_2},\\
	h_2&={\rm e}^{\eta_1}+{\rm e}^{\eta_2}+a_{13}a_{14}a_{34}{\rm e}^{\xi_1+\eta_1+\eta_2}+a_{23}a_{24}a_{34}{\rm e}^{\xi_2+\eta_1+\eta_2},\\
	f_2&=1+a_{13}{\rm e}^{\xi_1+\eta_1}+a_{14}{\rm e}^{\xi_1+\eta_2}+a_{23}{\rm e}^{\xi_2+\eta_1}+a_{24}{\rm e}^{\xi_2+\eta_2}+a_{12}a_{13}a_{14}a_{23}a_{24}a_{34}{\rm e}^{\xi_1+\xi_2+\eta_1+\eta_2}
	\end{split}
	\end{eqnarray}

where
\begin{eqnarray*}
	&&a_{13}=-\frac{1}{(k_1+l_1)^2},~a_{14}=-\frac{1}{(k_1+l_2)^2},~a_{23}=-\frac{1}{(k_2+l_1)^2},~a_{24}=-\frac{1}{(k_2+l_2)^2},\\&&a_{12}=-(k_1-k_2)^2,~ a_{34}=-(l_1-l_2)^2, ~
	\xi_i=k_ix+{\rm i}k_i^2t+\xi_{i0}, ~\eta_i=l_ix-{\rm i}l_i^2t+\eta_{i0}, i=1,2.
\end{eqnarray*}

For the	three-soliton solution, the expressions are 

\begin{eqnarray*}
	g_3&=&\sum_{1\leq i\leq 3}{\rm e}^{\xi_i}+\sum_{1\leq i<j\leq 3,1\leq p\leq 3} A_{ijp}{\rm e}^{\xi_i+\xi_j+\eta_p}+\sum_{1\leq p<q\leq 3} B_{pq}{\rm e}^{\xi_1+\xi_2+\xi_3+\eta_p+\eta_q},\\
	h_3&=&\sum_{1\leq p\leq 3}{\rm e}^{\eta_p}+\sum_{1\leq p<q\leq 3,1\leq i\leq 3} C_{pqi}{\rm e}^{\eta_p+\eta_q+\xi_i}+\sum_{1\leq i<j\leq 3} D_{ij}{\rm e}^{\eta_1+\eta_2+\eta_3+\xi_i+\xi_j},\\
	f_3&=&1+\sum_{1\leq i,p\leq 3} E_{ip}{\rm e}^{\xi_i+\eta_p}
	+\sum_{1\leq i<j\leq 3,1\leq p<q\leq 3} F_{ijpq}{\rm e}^{\xi_i+\xi_j+\eta_p+\eta_q}
	+G{\rm e}^{\xi_1+\xi_2+\xi_3+\eta_1+\eta_2+\eta_3}.
\end{eqnarray*}
The coefficients are defined as
\begin{eqnarray*}
	&&A_{ijp}=a_{ij}a_{ip}a_{jp},~ B_{pq}=a_{pq}\prod_{1\leq j<j\leq 3} a_{ij}\prod_{1\leq i\leq 3} a_{ip} \prod_{1\leq i\leq 3} a_{iq},\\
	&&C_{pqi}=a_{pq}a_{pi}a_{qi}, ~D_{ij}=a_{ij}\prod_{1\leq p<q\leq 3}a_{pq}\prod_{1\leq p\leq 3} a_{pi} \prod_{1\leq p\leq 3} a_{pj},\\
	&&E_{ip}=a_{ip},~ F_{ijpq}=\prod_{\text{all pairs}}a_{mn},  ~G=\prod_{1\leq i,p\leq 3} a_{ip}\prod_{1\leq i<j\leq 3}a_{ij}\prod_{1\leq p<q\leq 3}a_{pq}	
\end{eqnarray*}
with
\begin{eqnarray*}
	&&a_{ij}=-(k_i-k_j)^2,a_{pq}=-(l_p-l_q)^2, a_{ip}=a_{pi}=-\frac{1}{(k_i+l_p)^2},a_{jp}=a_{pj}=-\frac{1}{(k_j+l_p)^2}, \\&&  a_{iq}=a_{qi}=-\frac{1}{(k_i+l_q)^2}, ~\xi_i=k_ix+{\rm i}k_i^2t+\xi_{i0}, ~\eta_i=l_ix-{\rm i}l_i^2t+\eta_{i0}, i=1,2,3.
\end{eqnarray*}
In the given expression, the term ``all pairs" denotes the complete set of possible pairwise combinations of the indices \(i, j, p, q\), specifically \((i,j), (i,p), (i,q), (j,p), (j,q), (p,q)\).

The representations \eqref{nsoliton}
	satisfy the AKNS system identically and yield \(n\)-soliton manifolds parametrized by \(2n\) arbitrary  wave numbers and \(2n\) independent phase constants. This type formula makes the factorization of multi-soliton interactions transparent and is particularly suited for analyzing the asymptotic phase-shifts arising upon pairwise collisions.

\section{$K$- and $\tau$-symmetries of the AKNS system}

The AKNS system \eqref{akns} possesses an infinite-dimensional symmetry algebra encoded by two distinguished hierarchies of flows: the \( K \)-symmetries (isospectral deformations) and the \( \tau \)-symmetries (non-isospectral deformations). These symmetries are generated recursively by a common integro-differential recursion operator \( L \), whose action on an initial seed vector field produces the entire commuting vector fields.

More precisely, the \( m \)-th \( K \)-symmetry is defined as  \cite{ybib21}
\begin{eqnarray}
\boldsymbol{K}_{m}=\begin{pmatrix}
K_{m1} \vspace{0.2cm}\\
K_{m2}
\end{pmatrix}=L^m
\begin{pmatrix}
-{\rm i}q \vspace{0.2cm}\\
{\rm i}r 
\end{pmatrix},
\end{eqnarray}
while the corresponding \( \tau \)-symmetry is 
\begin{eqnarray}
\boldsymbol{\tau}_{m}=
\begin{pmatrix}
\tau_{m1} \vspace{0.2cm}\\
\tau_{m2}
\end{pmatrix}=L^m
\begin{pmatrix}
{\rm i}qx-2q_xt \vspace{0.2cm}\\
-{\rm i}rx-2r_xt 
\end{pmatrix},
\quad m=0,1,2,\dots,
\end{eqnarray}
where  the recursion operator
\[
L=-{\rm i}\begin{pmatrix}
-\partial_x+2q\partial_x^{-1}r  &~ 2q\partial_x^{-1}q \vspace{0.2cm}\\
-2r\partial_x^{-1}r &~ \partial_x-2r\partial_x^{-1}q 
\end{pmatrix}
\]
maps symmetries into symmetries and encodes the bi-Hamiltonian structure of the AKNS hierarchy.
The lowest members of the \( K \)-symmetry are
\begin{eqnarray*}
	&&
	\boldsymbol{K}_{1}=\begin{pmatrix}
		K_{11} \vspace{0.2cm}\\
		K_{12}
	\end{pmatrix}=\begin{pmatrix}
		q_x \vspace{0.2cm}\\
		r_x 
	\end{pmatrix},\\&&\boldsymbol{K}_{2}=
	\begin{pmatrix}
		K_{21} \vspace{0.2cm}\\
		K_{22}
	\end{pmatrix}={-\rm i}\begin{pmatrix}
		-q_{xx}+2q^2r \vspace{0.2cm}\\
		r_{xx}-2r^2q 
	\end{pmatrix}=\begin{pmatrix}
		q_t \vspace{0.2cm}\\
		r_t 
	\end{pmatrix},\\&&\boldsymbol{K}_{3}=
	\begin{pmatrix}
		K_{31} \vspace{0.2cm}\\
		K_{32}
	\end{pmatrix}=\begin{pmatrix}
		-q_{xxx}+6rqq_x\vspace{0.2cm}\\
		-r_{xxx}+6qrr_x
	\end{pmatrix}
\end{eqnarray*}
and
\begin{eqnarray*}\boldsymbol{K}_{4}=
	\begin{pmatrix}
		K_{41} \vspace{0.2cm}\\
		K_{42}
	\end{pmatrix}={-\rm i}\begin{pmatrix}
		q_{4x}-2q^2r_{xx}-8rqq_{xx}-4qr_xq_x-6rq_x^2+6q^3r^2\vspace{0.2cm}\\
		-r_{4x}+8qrr_{xx}+2r^2q_{xx}+4rq_xr_x+6qr_x^2-6r^3q^2
	\end{pmatrix},
\end{eqnarray*}
where \( \boldsymbol{K}_{2} \) coincides with the AKNS flow itself. Analogously, the first few \( \tau \)-symmetries read  
	\begin{eqnarray*}
	&&\boldsymbol{\tau}_{0}=
	\begin{pmatrix}
		\tau_{01} \vspace{0.2cm}\\
		\tau_{02}
	\end{pmatrix}=\begin{pmatrix}
		{\rm i}qx-2q_xt \vspace{0.2cm}\\
		-{\rm i}rx-2r_xt
	\end{pmatrix},
	\\&&\boldsymbol{\tau}_{1}=
	\begin{pmatrix}
		\tau_{11} \vspace{0.2cm}\\
		\tau_{12}
	\end{pmatrix}=\begin{pmatrix}
		2{\rm i}(2q^2rt-q_{xx}t)-q_xx-q \vspace{0.2cm}\\
		2{\rm i}(r_{xx}t-2qr^2t)-r_xx-r 
	\end{pmatrix},
	\\&&\boldsymbol{\tau}_{2}=
	\begin{pmatrix}
		\tau_{21} \vspace{0.2cm}\\
		\tau_{22}
	\end{pmatrix}=\begin{pmatrix}
		{\rm i}(2q^2rx-q_{xx}x+2q\partial^{-1}_x(qr)-2q_x)+2q_{xxx}t-12qq_xrt \vspace{0.2cm}\\
		{\rm i}(r_{xx}x-2qr^2x-2r\partial^{-1}_x(qr)+2r_x)+2r_{xxx}t-12qrr_xt 
	\end{pmatrix},
	\end{eqnarray*}
and
\begin{eqnarray*}\boldsymbol{\tau}_{3}=
	\begin{pmatrix}
		\vspace{-0.1cm}\\
		\tau_{31} \vspace{0.65cm}\\
		\tau_{32}\vspace{0.42cm}
	\end{pmatrix}=\begin{pmatrix}
		2{\rm i}(q_{4x}t-8qq_{xx}rt-2q^2r_{xx}t-4qq_xr_xt-6q_x^2rt+6q^3r^2t)+\\q_{xxx}x+3q_{xx}-2q^2r-6qq_xrx-4q\partial^{-1}_x(q_xr)-2q_x\partial^{-1}_x(qr) \vspace{0.4cm}\\
		2{\rm i}(8qrr_{xx}t+2q_{xx}r^2t-r_{4x}t+4q_xrr_xt+6qr_x^2t-6q^2r^3t)+\\
		r_{xxx}x+3r_{xx}-6qr^2-6qrr_xx+4r\partial^{-1}_x(q_xr)-2r_x\partial^{-1}_x(qr)
	\end{pmatrix}.
\end{eqnarray*}
Each vector field \( \boldsymbol{K}_{m} \) or \( \boldsymbol{\tau}_{m} \) is an infinitesimal generator of a one-parameter group of transformations leaving the AKNS system invariant, equivalently, they solve the linearized AKNS system  
		\begin{eqnarray}\label{sym}
		\begin{pmatrix}
		\sigma_{1} \vspace{0.2cm}\\
		\sigma_{2}
		\end{pmatrix}_t
		={\rm i}\begin{pmatrix}
		\sigma_{1xx}-2q^2\sigma_2-4qr\sigma_1 \vspace{0.2cm}\\
		-\sigma_{2xx}+2r^2\sigma_1+4qr\sigma_2 
		\end{pmatrix}.
		\end{eqnarray}
Consequently, the totality of \( K \)- and \( \tau \)-symmetries constitutes an infinite-dimensional Lie algebra that encodes the full integrability of the AKNS system, governs the structure of its soliton manifolds, and underlies the construction of  multi-soliton solutions via symmetry reduction and Darboux–B\"acklund transformations.

\section{Finite-dimensional reduction of $K$-symmetries on $n$-soliton solution \eqref{fnsolution}}

For the \( n \)-soliton solution \eqref{fnsolution}, the set of physical symmetries arising from translational invariance of individual soliton centers is encoded by the \( 2n \) parameter derivatives
\begin{eqnarray*}
q_{n,\xi_{i0}}=\left(\frac{Q_n}{M_n}\right)_{\xi_{i0}},\quad r_{n,\xi_{i0}}=\left(-\frac{R_n}{M_n}\right)_{\xi_{i0}},\quad i=1,...,n.
\end{eqnarray*} 
Each of which generates a translation of the centre of the \(i\)-th soliton.
These  translation symmetries constitute a canonical basis for the tangent space to the \(n\)-soliton stratum inside the full solution variety of the AKNS system.

One may legitimately ask whether the infinite hierarchies of generalized symmetries \( \{\boldsymbol{K}_m,\boldsymbol{\tau}_m\}_{m=0}^{\infty} \) generated by the AKNS recursion operator \( L \) collapse to the finite-dimensional span of the above physical symmetries when restricted to the \( n \)-soliton background \eqref{fnsolution}. The answer is affirmative for the \( K \)-symmetry and negative for the \( \tau \)-symmetry.

Indeed, restricting the $K$-symmetry to the $n$-soliton solution, one finds that each member reduces to a universal linear combination of the elementary shift derivatives, as follows
\begin{eqnarray*}
	&&K_{11}|_{q=q_n,r=r_n}=q_x|_{q=q_n,r=r_n}=\sum_{i=1}^nk_iq_{n,\xi_{i0}},\\
	&&K_{12}|_{q=q_n,r=r_n}=r_x|_{q=q_n,r=r_n}=\sum_{i=1}^nk_ir_{n,\xi_{i0}},\\
	&&K_{21}|_{q=q_n,r=r_n}=Lq_x|_{q=q_n,r=r_n}=-{\rm i}\sum_{i=1}^nk_i^2q_{n,\xi_{i0}},\\
	&&K_{22}|_{q=q_n,r=r_n}=Lr_x|_{q=q_n,r=r_n}=-{\rm i}\sum_{i=1}^nk_i^2r_{n,\xi_{i0}},\\
	&&K_{31}|_{q=q_n,r=r_n}=L^2q_x|_{q=q_n,r=r_n}=-\sum_{i=1}^nk_i^3q_{n,\xi_{i0}},\\
	&&K_{32}|_{q=q_n,r=r_n}=L^2r_x|_{q=q_n,r=r_n}=-\sum_{i=1}^nk_i^3r_{n,\xi_{i0}},\\
	&&\qquad \vdots\\
	&&K_{j1}|_{q=q_n,r=r_n}=L^{j-1}q_x|_{q=q_n,r=r_n}=({\rm -i})^{j-1}\sum_{i=1}^nk_i^jq_{n,\xi_{i0}},\\
	&&K_{j2}|_{q=q_n,r=r_n}=L^{j-1}r_x|_{q=q_n,r=r_n}=({\rm -i})^{j-1}\sum_{i=1}^nk_i^jr_{n,\xi_{i0}},~j=1,2,...,\infty.
\end{eqnarray*}
Hence every higher \( K \)-symmetry is a linear combination of the elementary center-translation symmetries with spectral coefficients \( (-\mathrm{i})^{j-1}k_i^j \). Consequently, the infinite set \( \{\boldsymbol{K}_j\} \)  collapses, upon restriction to the \(n\)-soliton background, to the  \( 2n \)-dimensional vector space.  More precisely,
For \( j\ge n+1 \) the vectors \( \boldsymbol{K}_j \) are linearly dependent on the first \( n \) flows, explicitly,
\begin{eqnarray*}
	&&K_{j1}|_{q=q_n,r=r_n}=L^{j-1}q_x|_{q=q_n,r=r_n}=({-\rm i})^{j-1}\sum_{i=1}^nk_i^j\left.\frac{\Delta_{M_i}}{\Delta_{M}}\right|_{q=q_n,r=r_n},\\
	&&K_{j2}|_{q=q_n,r=r_n}=L^{j-1}r_x|_{q=q_n,r=r_n}=({-\rm i})^{j-1}\sum_{i=1}^nk_i^j\left.\frac{\Delta_{N_i}}{\Delta_{M}}\right|_{q=q_n,r=r_n},\\&&j=n+1,\ n+2,\ \ldots, \ \infty.
\end{eqnarray*}
Here, $\Delta_{M}$ is utilized to denote the determinant of the $2n \times 2n$ matrix $ M $, which is formally defined as $ \Delta_{M} \equiv \det(M) $. The matrix $ M $ is constituted as
\begin{eqnarray}
M = \begin{pmatrix}
k_1  & \cdots &k_i&\cdots & k_n 
\\
{-\rm i}k_1^2  & \cdots &{-\rm i}k_i^2&\cdots &\ {-\rm i}k_n^2 
\\
-k_1^3  & \cdots &-k_i^3&\cdots& -k_n^3
\\
\vdots &  \ddots & \vdots & \ddots & \vdots \\
{(-\rm i)}^{n-1}k_1^{n}& \cdots &{(-\rm i)}^{n-1}k_i^{n}&\cdots& {(-\rm i)}^{n-1}k_n^{n}\\
\end{pmatrix}.
\end{eqnarray} $\Delta_{M_i},$ and $\Delta_{N_i}, $ represent the determinant of the $ n \times n $ matrix $M_i$ and $N_i$, by replacing the \(i\)-th column with the vectors \((K_{11},\dots,K_{n1})^{\mathsf{T}}\) and \((K_{12},\dots,K_{n2})^{\mathsf{T}}\), respectively,
\begin{eqnarray*}\label{Mi}
	M_{i} = \begin{pmatrix}
		k_1  & \cdots &K_{11}&\cdots & k_n 
		\\
		{-\rm i}k_1^2  & \cdots &K_{21}&\cdots &\ {-\rm i}k_n^2  
		\\
		-k_1^3  & \cdots &K_{31}&\cdots& -k_n^3
		\\
		\vdots &  \ddots & \vdots & \ddots & \vdots   \\
		{(-\rm i)}^{n-1}k_1^{n}& \cdots &K_{n1}&\cdots& {(-\rm i)}^{n-1}k_n^{n}\\
	\end{pmatrix},
\end{eqnarray*}
\begin{eqnarray*}\label{N1}
	N_{i} = \begin{pmatrix}
		k_1  & \cdots &K_{12}&\cdots & k_n 
		\\
		{-\rm i}k_1^2  & \cdots &K_{22}&\cdots &\ {-\rm i}k_n^2 
		\\
		-k_1^3  & \cdots &K_{32}&\cdots& -k_n^3
		\\
		\vdots &  \ddots & \vdots & \ddots & \vdots   \\
		{(-\rm i)}^{n-1}k_1^{n}& \cdots &K_{n2}&\cdots& {(-\rm i)}^{n-1}k_n^{n}\\
	\end{pmatrix}.
\end{eqnarray*}
For the lowest values of \(n\), the explicit expressions read as follows.
When \(n=1\), the solitary-wave reduction leaves only one independent generator. Every higher flow is merely a spectral rescaling of the basic translation symmetry
\begin{eqnarray*}
	&&K_{j1}=({-\rm i}k_1)^{j-1}K_{11},\\
	&&K_{j2}=({-\rm i}k_2)^{j-1}K_{12},~j\geq 2.
\end{eqnarray*}
For the two-soliton sector (\(n=2\)), all subsequent ones are rigidly determined by the first two symmetries
\begin{eqnarray*}
	&&K_{j1}=\frac{k_1({-\rm i}k_2)^{j-1}-k_2({-\rm i}k_1)^{j-1}}{k_1-k_2}K_{11}-\frac{{\rm i}[({-\rm i}k_2)^{j-1}-({-\rm i}k_1)^{j-1}]}{k_1-k_2}K_{21},\\
	&&K_{j2}=\frac{k_1({-\rm i}k_2)^{j-1}-k_2({-\rm i}k_1)^{j-1}}{k_1-k_2}K_{12}-\frac{{\rm i}[({-\rm i}k_2)^{j-1}-({-\rm i}k_1)^{j-1}]}{k_1-k_2}K_{22},~j\geq 3.
\end{eqnarray*}
In the three-soliton case (\(n=3\)), the first three generators form a basis. Every higher symmetry is expressed as the linear combination
\begin{eqnarray*}
	&&K_{j1}=\frac{A}{D}K_{11}+\frac{B}{D}K_{21}+\frac{C}{D}K_{31},\\
	&&K_{j2}=\frac{A}{D}K_{12}+\frac{B}{D}K_{22}+\frac{C}{D}K_{32}, ~j\geq 4
\end{eqnarray*}
with coefficient functions
\begin{eqnarray*}
&&A=(k_2-k_3)k_2k_3({-\rm i}k_1)^{j-1}+(k_3-k_1)k_1k_3({-\rm i}k_2)^{j-1}+(k_1-k_2)k_1k_2({-\rm i}k_3)^{j-1},\\
&&B={\rm i}[(k_3^2-k_2^2)({-\rm i}k_1)^{j-1}+(k_1^2-k_3^2)({-\rm i}k_2)^{j-1}+(k_2^2-k_1^2)({-\rm i}k_3)^{j-1}],\\
&&C=(k_3-k_2)({-\rm i}k_1)^{j-1}+(k_1-k_3)({-\rm i}k_2)^{j-1}+(k_2-k_1)({-\rm i}k_3)^{j-1},\\
&&D=(k_1-k_2)(k_1-k_3)(k_2-k_3).
\end{eqnarray*}
Thus the entire \( K \)-symmetry is finitely generated by the physical center-translation symmetries on the \( n \)-soliton background.

By contrast, the \(\tau\)-symmetries do not reduce to linear combinations of the center-translation vectors \(\{q_{n,\xi_{i0}},r_{n,\xi_{i0}}\}\) and no algebraic identity relates the \(\boldsymbol{\tau}_{m}\) hierarchy to the \(n\)-soliton \eqref{fnsolution} background.	
		
\section{Finite-dimensional reduction of $K$-symmetries on $n$-soliton solution \eqref{nsoliton}
}
For the  \(n\)-soliton solution \eqref{nsoliton}, the \(4n\) elementary infinitesimal translations are generated by the phase derivatives  

\begin{eqnarray*}
q_{n,\xi_{i0}}=\left(\frac{g_n}{f_n}\right)_{\xi_{i0}},\quad 
r_{n,\xi_{i0}}=\left(\frac{h_n}{f_n}\right)_{\xi_{i0}},\quad q_{n,\eta_{i0}}=\left(\frac{g_n}{f_n}\right)_{\eta_{i0}},\quad
r_{n,\eta_{i0}}=\left(\frac{h_n}{f_n}\right)_{\eta_{i0}},\quad i=1,\dots,n.
\end{eqnarray*}
Evaluating the \(K\)-symmetry on this background yields the reconstruction formulas
		\begin{eqnarray*}
			&&K_{11}|_{q=q_n,r=r_n}=q_x|_{q=q_n,r=r_n}=\sum_{i=1}^nk_iq_{n,\xi_{i0}}+\sum_{i=1}^nl_iq_{n,\eta_{i0}},\\
			&&K_{12}|_{q=q_n,r=r_n}=r_x|_{q=q_n,r=r_n}=\sum_{i=1}^nk_ir_{n,\xi_{i0}}+\sum_{i=1}^nl_ir_{n,\eta_{i0}},\\
			&&K_{21}|_{q=q_n,r=r_n}=Lq_x|_{q=q_n,r=r_n}={\rm i}\sum_{i=1}^nk_i^2q_{n,\xi_{i0}}-{\rm i}\sum_{i=1}^nl_i^2q_{n,\eta_{i0}},\\
			&&K_{22}|_{q=q_n,r=r_n}=Lr_x|_{q=q_n,r=r_n}={\rm i}\sum_{i=1}^nk_i^2r_{n,\xi_{i0}}-{\rm i}\sum_{i=1}^nl_i^2r_{n,\eta_{i0}},\\
			&&K_{31}|_{q=q_n,r=r_n}=L^2q_x|_{q=q_n,r=r_n}=-\sum_{i=1}^nk_i^3q_{n,\xi_{i0}}-\sum_{i=1}^nl_i^3q_{n,\eta_{i0}},\\
			&&K_{32}|_{q=q_n,r=r_n}=L^2r_x|_{q=q_n,r=r_n}=-\sum_{i=1}^nk_i^3r_{n,\xi_{i0}}-\sum_{i=1}^nl_i^3r_{n,\eta_{i0}},\\
			&&\qquad \vdots\\
			&&K_{j1}|_{q=q_n,r=r_n}=L^{j-1}q_x|_{q=q_n,r=r_n}={\rm i}^{j-1}\sum_{i=1}^nk_i^jq_{n,\xi_{i0}}+{(-\rm i)}^{j-1}\sum_{i=1}^nl_i^jq_{n,\eta_{i0}},\\
			&&K_{j2}|_{q=q_n,r=r_n}=L^{j-1}r_x|_{q=q_n,r=r_n}={\rm i}^{j-1}\sum_{i=1}^nk_i^jr_{n,\xi_{i0}}+{(-\rm i)}^{j-1}\sum_{i=1}^nl_i^jr_{n,\eta_{i0}},~j=1,2,...,\infty.
		\end{eqnarray*}
		Consequently the infinite sequence \(\{K_{j}\}_{j=1}^{\infty}\) collapses to a \(4n\)-dimensional vector space. Every higher symmetry with \(j\geq 2n+1\) is expressed as the linear combination
		\begin{eqnarray*}
			&&K_{j1}|_{q=q_n,r=r_n}=L^{j-1}q_x|_{q=q_n,r=r_n}={\rm i}^{j-1}\sum_{i=1}^nk_i^j\left.\frac{\Delta_{M_i}}{\Delta_{M}}\right|_{q=q_n,r=r_n}+{(-\rm i)}^{j-1}\sum_{i=1}^nl_i^j\left.\frac{\Delta_{M_{ni}}}{\Delta_{M}}\right|_{q=q_n,r=r_n},\\
			&&K_{j2}|_{q=q_n,r=r_n}=L^{j-1}r_x|_{q=q_n,r=r_n}={\rm i}^{j-1}\sum_{i=1}^nk_i^j\left.\frac{\Delta_{N_i}}{\Delta_{M}}\right|_{q=q_n,r=r_n}+{(-\rm i)}^{j-1}\sum_{i=1}^nl_i^j\left.\frac{\Delta_{N_{ni}}}{\Delta_{M}}\right|_{q=q_n,r=r_n},
		\end{eqnarray*}
	where \(\Delta_{M}\) is the determinant of the \(2n \times 2n\)   matrix
		\begin{eqnarray*}
		M = \begin{pmatrix}
		k_1  & \cdots &k_i&\cdots & k_n & l_1 &  \cdots& l_i  & \cdots & l_n 
		\\
		{\rm i}k_1^2  & \cdots &{\rm i}k_i^2&\cdots &\ {\rm i}k_n^2&-{\rm i}l_1^2 & \cdots &-{\rm i}l_i^2&\cdots& -{\rm i}l_{n}^2  
		\\
		-k_1^3  & \cdots &-k_i^3&\cdots& -k_n^3&-l_1^3 & \cdots & -l_{i}^3  &\cdots &-l_{n}^3
		\\
		\vdots &  \ddots & \vdots & \ddots & \vdots &\vdots & \ddots&\vdots &  \ddots & \vdots  \\
		{\rm i}^{2n-1}k_1^{2n}& \cdots &{\rm i}^{2n-1}k_i^{2n}&\cdots& {\rm i}^{2n-1}k_n^{2n}&({-\rm i})^{2n-1} l_1^{2n} & \cdots & (-{\rm i})^{2n-1}l_{i}^{2n} &\cdots &(-{\rm i})^{2n-1}l_{n}^{2n}\\
		\end{pmatrix}.
		\end{eqnarray*}
		 $\Delta_{M_i}$ and $ \Delta_{M_{ni}}$ represent the determinant of the $ 2n \times 2n $ matrix $M_i$ and $M_{ni}$ by replacing the appropriate column with the vectors \((K_{11},\dots ,K_{2n,1})^{\top}\)
		\begin{eqnarray*}
			M_{i} = \begin{pmatrix}
				k_1  & \cdots &K_{11}&\cdots & k_n & l_1 &  \cdots& l_i  & \cdots & l_n 
				\\
				{\rm i}k_1^2  & \cdots &K_{21}&\cdots &\ {\rm i}k_n^2&-{\rm i}l_1^2 & \cdots &-{\rm i}l_i^2&\cdots& -{\rm i}l_{n}^2  
				\\
				-k_1^3  & \cdots &K_{31}&\cdots& -k_n^3&-l_1^3 & \cdots & -l_{i}^3  &\cdots &-l_{n}^3
				\\
				\vdots &  \ddots & \vdots & \ddots & \vdots &\vdots & \ddots&\vdots &  \ddots & \vdots  \\
				{\rm i}^{2n-1}k_1^{2n}& \cdots &K_{2n,1}&\cdots& {\rm i}^{2n-1}k_n^{2n}&({-\rm i})^{2n-1} l_1^{2n} & \cdots & (-{\rm i})^{2n-1}l_{i}^{2n} &\cdots &(-{\rm i})^{2n-1}l_{n}^{2n}\\
			\end{pmatrix},
		\end{eqnarray*}
		
		\begin{eqnarray*}
			M_{ni} = \begin{pmatrix}
				k_1  & \cdots &k_i&\cdots & k_n & l_1 &  \cdots& K_{11}  & \cdots & l_n 
				\\
				{\rm i}k_1^2  & \cdots &{\rm i}k_i^2&\cdots &\ {\rm i}k_n^2&-{\rm i}l_1^2 & \cdots &K_{21}&\cdots& -{\rm i}l_{n}^2  
				\\
				-k_1^3  & \cdots &-k_i^3&\cdots& -k_n^3&-l_1^3 & \cdots & K_{31} &\cdots &-l_{n}^3
				\\
				\vdots &  \ddots & \vdots & \ddots & \vdots &\vdots & \ddots&\vdots &  \ddots & \vdots  \\
				{\rm i}^{2n-1}k_1^{2n}& \cdots &{\rm i}^{2n-1}k_i^{2n}&\cdots& {\rm i}^{2n-1}k_n^{2n}&({-\rm i})^{2n-1} l_1^{2n} & \cdots & K_{2n,1} &\cdots &(-{\rm i})^{2n-1}l_{n}^{2n}\\
			\end{pmatrix}.
		\end{eqnarray*}
		$\Delta_{N_i}$ and $\Delta_{N_{ni}}$  represent the determinant of the $ 2n \times 2n $ matrix $N_i$ and $N_{ni}$ by replacing the corresponding column with the vectors  \((K_{12},\dots ,K_{2n,2})^{\top}\), respectively,
		\begin{eqnarray*}
			N_{i} = \begin{pmatrix}
				k_1  & \cdots &K_{12}&\cdots & k_n & l_1 &  \cdots& l_i  & \cdots & l_n 
				\\
				{\rm i}k_1^2  & \cdots &K_{22}&\cdots &\ {\rm i}k_n^2&-{\rm i}l_1^2 & \cdots &-{\rm i}l_i^2&\cdots& -{\rm i}l_{n}^2  
				\\
				-k_1^3  & \cdots &K_{32}&\cdots& -k_n^3&-l_1^3 & \cdots & -l_{i}^3  &\cdots &-l_{n}^3
				\\
				\vdots &  \ddots & \vdots & \ddots & \vdots &\vdots & \ddots&\vdots &  \ddots & \vdots  \\
				{\rm i}^{2n-1}k_1^{2n}& \cdots &K_{2n,2}&\cdots& {\rm i}^{2n-1}k_n^{2n}&({-\rm i})^{2n-1} l_1^{2n} & \cdots & (-{\rm i})^{2n-1}l_{i}^{2n} &\cdots &(-{\rm i})^{2n-1}l_{n}^{2n}\\
			\end{pmatrix},
		\end{eqnarray*}
		
		\begin{eqnarray*}\label{N2}
			N_{ni} = \begin{pmatrix}
				k_1  & \cdots &k_i&\cdots & k_n & l_1 &  \cdots& K_{12}  & \cdots & l_n 
				\\
				{\rm i}k_1^2  & \cdots &{\rm i}k_i^2&\cdots &\ {\rm i}k_n^2&-{\rm i}l_1^2 & \cdots &K_{22}&\cdots& -{\rm i}l_{n}^2  
				\\
				-k_1^3  & \cdots &-k_i^3&\cdots& -k_n^3&-l_1^3 & \cdots & K_{32} &\cdots &-l_{n}^3
				\\
				\vdots &  \ddots & \vdots & \ddots & \vdots &\vdots & \ddots&\vdots &  \ddots & \vdots  \\
				{\rm i}^{2n-1}k_1^{2n}& \cdots &{\rm i}^{2n-1}k_i^{2n}&\cdots& {\rm i}^{2n-1}k_n^{2n}&({-\rm i})^{2n-1} l_1^{2n} & \cdots & K_{2n,2} &\cdots &(-{\rm i})^{2n-1}l_{n}^{2n}\\
			\end{pmatrix}.
		\end{eqnarray*}
		Closed-form expressions for the lowest values of \(n\) are as follows. When \( n=1 \), every higher symmetry is a linear combination of the two elementary generators
		\begin{eqnarray*}
			&&K_{j1}=\left(\frac{({\rm i}k_1)^{j-1}l_1+k_1(-{\rm i}l_1)^{j-1}}{k_1+l_1}\right)K_{11}+\left(\frac{{\rm i}(-{\rm i}l_1)^{j-1}-{\rm i}({\rm i}k_1)^{j-1}}{k_1+l_1}\right)K_{21},\\
			&&K_{j2}=\left(\frac{({\rm i}k_1)^{j-1}l_1+k_1(-{\rm i}l_1)^{j-1}}{k_1+l_1}\right)K_{12}+\left(\frac{{\rm i}(-{\rm i}l_1)^{j-1}-{\rm i}({\rm i}k_1)^{j-1}}{k_1+l_1}\right)K_{22},~j\geq 3.
		\end{eqnarray*}
		
		For \( n=2 \), the first four symmetries span the space and all subsequent ones are reconstructed as
		\begin{eqnarray*}
			&&	K_{j1}=A K_{11}
			+B K_{21}
			-C K_{31}
			-D K_{41},\\
			&&	K_{j2}=A K_{12}
			+B K_{22}
			-C K_{32}
			-D K_{42}, ~j\geq 5
		\end{eqnarray*}
	with coefficients
		\begin{eqnarray*}
			A&=&\left(-\frac{(-{\rm i }l_1)^{i-1}k_1k_2l_2}{(k_2+l_1)(l_1-l_2)(k_1+l_1)}+\frac{(-{\rm i }l_2)^{i-1}k_1k_2l_1}{(k_2+l_2)(k_1+l_2)(l_1-l_2)}\right.\\&-&\left.\frac{({\rm i }k_1)^{i-1}k_2l_1l_2}{(k_1-k_2)(k_1+l_1)(k_1+l_2)}+\frac{({\rm i }k_2)^{i-1}k_1l_1l_2}{(k_1-k_2)(k_2+l_1)(k_2+l_2)}\right),\\
			B&=&{\rm i}\left(\frac{(-{\rm i }l_1)^{i-1}(k_1k_2-k_1l_2-k_2l_2)}{(k_2+l_1)(l_1-l_2)(k_1+l_1)}-\frac{(-{\rm i }l_2)^{i-1}(k_1k_2-k_1l_1-k_2l_1)}{(k_1+l_2)(k_2+l_2)(l_1-l_2)}\right.\\&+&\left.\frac{({\rm i }k_1)^{i-1}(k_2l_1+k_2l_2-l_1l_2)}{(k_1-k_2)(k_1+l_2)(k_1+l_1)}-\frac{({\rm i }k_2)^{i-1}(k_1l_1+k_1l_2-l_1l_2)}{(k_1-k_2)(k_2+l_1)(k_2+l_2)}\right),\\
			C&=&\left(\frac{(-{\rm i }l_1)^{i-1}(k_1+k_2-l_2)}{(k_1+l_1)(k_2+l_1)(l_1-l_2)}-\frac{(-{\rm i }l_2)^{i-1}(k_1+k_2-l_1)}{(k_1+l_2)(k_2+l_2)(l_1-l_2)}\right.\\&-&\left.\frac{({\rm i }k_1)^{i-1}(k_2-l_1-l_2)}{(k_1-k_2)(k_1+l_1)(k_1+l_2)}+\frac{({\rm i }k_2)^{i-1}(k_1-l_1-l_2)}{(k_1-k_2)(k_2+l_1)(k_2+l_2)}\right),\\
			D&=&{\rm i}\left(\frac{(-{\rm i }l_1)^{i-1}}{(k_1+l_1)(k_2+l_1)(l_1-l_2)}-\frac{(-{\rm i }l_2)^{i-1}}{(k_1+l_2)(k_2+l_2)(l_1-l_2)}\right.\\&-&\left.\frac{({\rm i }k_1)^{i-1}}{(k_1-k_2)(k_1+l_1)(k_1+l_2)}+\frac{({\rm i }k_2)^{i-1}}{(k_1-k_2)(k_2+l_1)(k_2+l_2)}\right).
		\end{eqnarray*}

\section {Finite-dimensional reduction of $\tau$-symmetries on one-soliton solution \eqref{1-s}}

		We now establish the precise linear combination of the $\tau$-symmetries in terms of the elementary wave-parameter derivatives on the one-soliton solution \eqref{1-s}. Let
		\begin{eqnarray*}
		q_{k_1}=\partial_{k_1}q,\quad q_{l_1}=\partial_{l_1}q,\quad q_{\xi_{10}}=\partial_{\xi_{10}}q,\quad q_{\eta_{10}}=\partial_{\eta_{10}}q,
		\end{eqnarray*} 
	and similarly for the conjugate component $r$.
A direct computation gives the first few $\tau$-symmetries
		\begin{eqnarray*}
			\tau_{01}&=&{\rm i}q_{k_1}-{\rm i}q_{l_1},\\
			\tau_{11}&=&-k_1q_{k_1}-l_1q_{l_1}-q_{\xi_{10}}-q_{\eta_{10}},\\
			\tau_{21}&=&-{\rm i}k_1^2q_{k_1}+{\rm i}l_1^2q_{l_1}+2{\rm i}l_1q_{\xi_{10}}-2{\rm i}k_1q_{\eta_{10}}\\
			&=&\frac{{\rm i}(l_1^2-k_1^2)}{k_1l_1}K_{11}+\frac{(k_1+l_1)}{k_1l_1}K_{21}-k_1l_1\tau_{01}+{\rm i}(k_1-l_1)\tau_{11},\\
			\tau_{31}&=&k_1^3q_{k_1}+l_1^3q_{l_1}-(k_1^2+2k_1l_1-2l_1^2)q_{\xi_{10}}+(2k_1^2-2k_1l_1-l_1^2)q_{\eta_{10}}\\
			&=&\frac{(k_1+l_1)(k_1^2-3k_1l_1+l_1^2)}{k_1l_1}K_{11}+\frac{{\rm i}(k_1^2-l_1^2)}{k_1l_1}K_{21}-{\rm i}(k_1-l_1)k_1l_1\tau_{01}-(k_1^2-k_1l_1+l_1^2)\tau_{11}.
		\end{eqnarray*}
The universal closed-form expression for $j\ge 2$ is
		\begin{eqnarray*}
		\tau_{j1}&=&-{\rm i }\left(\frac{ A_{j}}{k_1^2l_1^2(k_1+l_1)^2}\right)K_{11}+\left(\frac{B_{j}}{k_1^2l_1^2(k_1+l_1)^2}\right)K_{21}\\&+&\left(\frac{({\rm i}k_1)^jl_1+(-{\rm i}l_1)^jk_1}{k_1+l_1}\right)\tau_{01}-{\rm i}\left(\frac{({\rm i}k_1)^j-(-{\rm i} l_1)^j}{k_1+l_1}\right)\tau_{11}
	\end{eqnarray*}
with coefficient polynomials
	\begin{eqnarray*}
		A_{j}&=&((1-j)(-{\rm i} l_1)^j-({\rm i}k_1)^j)k_1^3l_1+((-{\rm i}l_1)^j+(j-1)({\rm i}k_1)^j)k_1l_1^3+j({\rm i}k_1)^jl_1^4-j(-{\rm i}l_1)^jk_1^4,\\
		B_{j}&=&((1-j)(-{\rm i} l_1)^j-({\rm i}k_1)^j)k_1^2l_1-((-{\rm i}l_1)^j+(j-1)({\rm i}k_1)^j)k_1l_1^2-j({\rm i}k_1)^jl_1^3-j(-{\rm i}l_1)^jk_1^3.
	\end{eqnarray*}

	An identical reconstruction holds for the conjugate component	
		\begin{eqnarray*}
			\tau_{02}&=&{\rm i}r_{k_1}-{\rm i}r_{l_1},\\
			\tau_{12}&=&-k_1r_{k_1}-l_1r_{l_1}-r_{\xi_{10}}-r_{\eta_{10}},\\
			\tau_{22}&=&-{\rm i}k_1^2r_{k_1}+{\rm i}l_1^2r_{l_1}+2{\rm i}l_1r_{\xi_{10}}-2{\rm i}k_1r_{\eta_{10}}\\
			&=&\frac{{\rm i}(l_1^2-k_1^2)}{k_1l_1}K_{12}+\frac{(k_1+l_1)}{k_1l_1}K_{22}-k_1l_1\tau_{02}+{\rm i}(k_1-l_1)\tau_{12},\\
			\tau_{32}&=&k_1^3r_{k_1}+l_1^3r_{l_1}-(k_1^2+2k_1l_1-2l_1^2)r_{\xi_{10}}+(2k_1^2-2k_1l_1-l_1^2)r_{\eta_{10}}\\
			&=&\frac{(k_1+l_1)(k_1^2-3k_1l_1+l_1^2)}{k_1l_1}K_{12}+\frac{{\rm i}(k_1^2-l_1^2)}{k_1l_1}K_{22}-{\rm i}(k_1-l_1)k_1l_1\tau_{02}-(k_1^2-k_1l_1+l_1^2)\tau_{12},
		\end{eqnarray*}
	and for \(j\ge 2\),
		\begin{eqnarray*}
			\tau_{j2}&=&-{\rm i }\left(\frac{ A_{j}}{k_1^2l_1^2(k_1+l_1)^2}\right)K_{12}+\left(\frac{B_{j}}{k_1^2l_1^2(k_1+l_1)^2}\right)K_{22}\\&+&\left(\frac{({\rm i}k_1)^jl_1+(-{\rm i}l_1)^jk_1}{k_1+l_1}\right)\tau_{02}-{\rm i}\left(\frac{({\rm i}k_1)^j-(-{\rm i} l_1)^j}{k_1+l_1}\right)\tau_{12}.
		\end{eqnarray*}
	Thus every $\tau$-symmetry is a linear combination of the two elementary $K$-symmetries $K_{11}, K_{21}$ and the two $\tau$-symmetries $T_{01}, T_{11}$, and similarly for the conjugate component. 
	
		The structural pattern observed for the one-soliton solution persists for the general \(n\)-soliton background. 
		Indeed, for an \(n\)-soliton solution endowed with \(2n\) independent spectral parameters \(\{k_i,l_i\}_{i=1}^n\) and the associated phase constants \(\{\xi_{i0},\eta_{i0}\}_{i=1}^n\), every \(\tau\)-symmetry can still be expressed as a  linear combination of a finite set of basis fields. 
		  Thus, on the generic \(n\)-soliton manifold, the entire $\tau$-symmetry is a finite-dimensional module over the field of rational functions in the spectral variables.

		  More generally, the preceding analysis extends mutatis mutandis to arbitrary multi-wave backgrounds carrying a discrete set of independent wave parameters together with their associated phase constants. 
		  On such manifolds the infinite hierarchy of $K$-symmetries collapses to a finite set generated by the  elementary translational vector fields. Every higher $K$-flow is reconstructed by a universal formula. A similar structural result holds for the $\tau$-symmetry. The full $\tau$-symmetry restricted to the multi-wave solution becomes a finite-dimensional vector space.

\section{Two-wave solution of the AKNS system via $K$-symmetries}
By systematically implementing the symmetry-constraint machinery of \cite{lou1}, we extract the two-wave solution of the AKNS system from its hierarchy of $K$-symmetries. Imposing the  ansatz
\begin{eqnarray*}
q=Q(\xi_{1},\xi_{2}),
\quad
r=R(\xi_{1},\xi_{2}),
\quad
\xi_{i}=k_{i}x+\omega_{i}t+\xi_{i0},\quad i=1,2
\end{eqnarray*}
with undetermined parameters \(k_{i},\omega_{i}\) and phase constants \(\xi_{i0}\), we substitute into the AKNS system \eqref{akns} and eliminate the temporal derivatives by virtue of \(\partial_{t}=\sum_{i}\omega_{i}\partial_{\xi_{i}}\) to obtain the reduced system
\begin{eqnarray*}
\sum_{i=1}^2\omega_{i}Q_{\xi_i}=
{\rm i}\sum_{i,j=1}^2k_ik_jQ_{\xi_i\xi_j}-2{\rm i}Q^2R,~~\sum_{i=1}^2\omega_{i}R_{\xi_i}=-{\rm i}\sum_{i,j=1}^2k_ik_jR_{\xi_i\xi_j}+2{\rm i}R^2Q.
\end{eqnarray*}
Inserting the ansatz into the  $K$-flows \(\boldsymbol{K}_{3}\) and \(\boldsymbol{K}_{4}\), and
setting
$\omega_i=-{\rm i}k_i^2, i=1,2,$ which is precisely the standard soliton dispersion relation of the AKNS system,
we get the two-soliton solution \eqref{2-sol}
\begin{eqnarray*}
	&&q=\frac{1+{\rm e^{\xi_1}}+{\rm e^{\xi_2}}}{k_1^2{\rm e^{\xi_1}}+k_2^2{\rm e^{\xi_2}}+(k_1-k_2)^2{\rm e^{\xi_1+\xi_2}}},\\
	&&r=-\frac{k_1^2k_2^2(k_1-k_2)^2{\rm e^{\xi_1+\xi_2}}}{k_1^2{\rm e^{\xi_1}}+k_2^2{\rm e^{\xi_2}}+(k_1-k_2)^2{\rm e^{\xi_1+\xi_2}}}.
\end{eqnarray*}	
	Its one-soliton reduction is obtained by switching off the aforementioned second phase	
\begin{eqnarray*}
q=\frac{1}{1+{\rm e}^{\xi_1}},\quad r=-\frac{k_1^2{\rm e}^{\xi_1}}{1+{\rm e}^{\xi_1}},\quad \xi_{1}=k_{1}x-\mathrm{i}k_{1}^{2}t+\xi_{10}.
\end{eqnarray*}
Adopting the alternative dispersion pair
\begin{eqnarray*}
\omega_1={\rm i}k_1^2, \quad \omega_2=-{\rm i}k_2^2,
\end{eqnarray*}
we recover the type-II one-soliton solution \eqref{1-s}
\begin{eqnarray*}
q=\frac{e^{\xi_1}}{1+e^{\xi_1+\xi_2}}, \quad r=-\frac{(k_1+k_2)^2e^{\xi_2}}{1+e^{\xi_1+\xi_2}}.
\end{eqnarray*}

Beyond these canonical cases, the symmetry-constraint algorithm generates some two-wave solutions by allowing some dispersion laws. Representative examples include:	\\
case 1: $\omega_1={\rm i}k_1^2,~ \omega_2={\rm i}k_2^2$
\begin{eqnarray*}
q=\frac{e^{\xi_1+\xi_2}}{e^{\xi_1}+e^{\xi_2}},\quad r=-\frac{(k_1-k_2)^2}{e^{\xi_1}+e^{\xi_2}}.
\end{eqnarray*}	
case 2: $\omega_1={\rm i}k_1^2, ~ \omega_2={\rm i}k_2^2,~ k_1=\frac{k_2}{2}$
\begin{eqnarray*}
	q=\frac{e^{\xi_1}+e^{\xi_2}+e^{\xi_1+\xi_2}}{1+e^{\xi_2}+4e^{\xi_1}},\quad r=-\frac{k_2^2}{1+e^{\xi_2}+4e^{\xi_1}}.
\end{eqnarray*}		
case 3:  	$\omega_1={\rm i}k_1^2,~ \omega_2=-{\rm i}k_2^2,~ k_1=-\frac{k_2}{2}$	
\begin{eqnarray*}
	q=\frac{1+e^{\xi_1}+e^{\xi_1+\xi_2}}{1+e^{\xi_2}+4e^{\xi_1+\xi_2}},\quad r=-\frac{k_2^2e^{\xi_2}}{1+e^{\xi_2}+4e^{\xi_1+\xi_2}}.
\end{eqnarray*}	
case 4: $\omega_1={\rm i}k_1^2$	 (arbitrary $\omega_2$)
\begin{eqnarray*}
q=-\frac{k_1^2(e^{\xi_1}-e^{\xi_1+\xi_2})}{1+e^{\xi_1}-e^{\xi_2}-e^{\xi_1+\xi_2}},\quad r=\frac{1+e^{\xi_2}}{1+e^{\xi_1}+e^{\xi_2}+e^{\xi_1+\xi_2}}.
\end{eqnarray*}				
case 5: $\omega_2=-{\rm i}k_2^2$  (arbitrary $\omega_1$)
\begin{eqnarray*}
q=\frac{1+e^{\xi_1}}{1+e^{\xi_1}+e^{\xi_2}+e^{\xi_1+\xi_2}},\quad r=-\frac{k_2^2e^{\xi_2}}{1+e^{\xi_2}}.
\end{eqnarray*}

Each of these closed-form expressions is an exact two-wave solution of the AKNS system and arises from a distinct choice of the dispersion relations imposed by the generalised $K$-symmetry constraints.

\section{Conclusions and discussions}

In this work, we have provided a characterization of the infinite-dimensional symmetry algebra of the AKNS system when the system is restricted to an $n$-soliton background.  For the $K$-symmetries, the collapse is complete and explicit. 
Each higher order $K$-symmetry is expressible as a  linear combination of the elementary center-translation symmetries, with coefficients determined by the wave parameters, so the entire $K$-hierarchy degenerates to a finite-dimensional module over the field of wave parameters. This reduction mirrors the finite-dimensional effective dynamics of solitons, linking symmetry algebra to observable wave behavior.

In sharp contrast, the $\tau$-symmetries do not admit any analogous decomposition in terms of center-translations on the $n$-soliton \eqref{fnsolution}. Nevertheless, once the background is enlarged to the multi-wave soliton \eqref{nsoliton}, which carries $4n$ independent wave parameters $\{k_i,l_i,\xi_{i0},\eta_{i0}\}$, the $\tau$-symmetries also become finite-dimensional.  
We have constructed a minimal basis of four symmetry vector fields, the two lowest $K$-flows together with the two primary $\tau$-flows on the one-soliton manifold. Every higher $\tau$-symmetry is then expressed as an explicit linear combination of these four generators, with coefficients determined solely by the wave parameters.

Beyond its conceptual value, the symmetry-constraint machinery yields reconstruction formulas that convert symbolic symmetry conditions into explicit multi-wave solutions, offering a new, algorithmic route to exact solutions of integrable systems.  The framework is universal and immediately transferable to any system endowed with a recursion operator and a rich solution space.
\\\\		
\noindent \bf Funding 
\rm The work was supported by the National Natural Science Foundations of China (Nos. 12301315,  12235007, 11975131), Zhejiang Provincial Natural Science Foundation of China No. LMS25A010002.
	\\\\	
\noindent \bf Author Contributions \rm X. Z. Hao contributed to the original draft of the manuscript and performed the formal analysis. S. Y. Lou provided the methodology, supervised the research, and reviewed and edited the manuscript.
	\\\\	
\noindent \bf Data Availability
		\rm No datasets were generated or analyzed during the current study.
\section*{\Large Declarations}
		\noindent \bf Conflict of interest \rm The authors declare no conflict of interest.
		
		\bibliographystyle{elsarticle-num}
		\bibliography{ref}

\end{document}